\documentclass[a4paper, 12pt]{article} 

\title{Inference for spatial processes using imperfect data from measurements and numerical simulations}
\author{Benjamin D. Youngman and David B. Stephenson}
\date{\today}
  
\usepackage{setspace}
\usepackage{amsmath}
\usepackage{graphicx}
\usepackage{dsfont}
\usepackage{bm}
\usepackage{natbib}
\usepackage{subfig}
\usepackage{hyperref}
\usepackage{mathtools}
\usepackage{gensymb}
\usepackage{setspace}

\newlength{\ntw}
\newlength{\nth}
\begin{document}

\maketitle

\section*{Abstract} \label{abstract}

We present a framework for inference for spatial processes that have actual values imperfectly represented by data. Environmental processes represented as spatial fields, either at fixed time points, or aggregated over fixed time periods, are studied. Data from both measurements and simulations performed by complex computer models are used to infer actual values of the spatial fields.

Methods from geostatistics and statistical emulation are used to explicitly capture discrepancies between a spatial field's actual and simulated values. A geostatistical model captures spatial discrepancy: the difference in spatial structure between simulated and actual values. An emulator represents the intensity discrepancy: the bias in simulated values of given intensity. Measurement error is also represented. Gaussian process priors represent each source of error, which gives an analytical expression for the posterior distribution for the actual spatial field.

Actual footprints for 50 European windstorms, which represent maximum wind gust speeds on a grid over a 72-hour period, are derived from wind gust speed measurements taken at stations across Europe and output simulated from a downscaled version of the Met Office Unified Model. The derived footprints have realistic spatial structure, and gust speeds closer to the measurements than originally simulated.

\section{Introduction} \label{intro}

Spatial fields are used to represent many different types of environmental process, such as maps of weather forecasts or radar images of past events. The environmental process may only be of partial interest; instead it may contribute towards estimating something else. One example is catastrophe models, which are used to provide loss estimates for natural hazards, such as earthquakes, hurricanes, floods and tsunamis; see \cite{grossi}. These combine data on natural hazard events, such as the intensity of the event at different locations, with data on property, such as its value and susceptibility to damage, to produce loss estimates. Reliable loss estimates require adequate hazard data. We study European windstorm footprints, which represent the maximum 3 second wind gust speed attained during an event over a 72-hour time period. Simulations of these can under-represent intensity \citep{xwspaper}, whereas measurements tend to be imprecise and spatially sparse. Resulting loss estimates from either may therefore be unreliable. We propose a framework that lets us explicitly consider how both types of data relate to actual values of a process under study, which allows us to make inference on its actual values, which cannot be observed. 

The accuracy of measurements on environmental processes can vary considerably between processes. Locations for which measurements are available are finite, and usually coincide with an irregularly-spread set of stations, which may not be located at a point of interest. Spatial field data overcome this. Numerical simulation models, such as variants of climate models, typically offer spatially complete data. We recognise that, however well calibrated or high in resolution, a numerical simulation model will always offer an imperfect representation of reality, that is, of the actual spatial field of the environmental process. This discrepancy between reality and a model's estimate must be recognised before reliable inferences about reality can be made. This work will use both measurements and numerical model simulations---and explicitly account for the deficiencies in each source of data---to derive probabilistic representations of actual spatial fields.

We assume that the spatial discrepancy between a simulated and actual spatial field takes a smooth form. We also assume that the intensity discrepancy, which is the mean bias in the environmental process for a given intensity, takes a smooth form. To accommodate these assumptions in a robust way we adopt Gaussian process priors. We combine covariance forms used in geostatistics and statistical emulation; see \cite{dig-rib} and \cite{oak} for respective overviews.  We can also readily assume within this approach that measurement errors, given actual values, are normally distributed and arise independently. Our assumptions result in a posterior distribution---with an analytical expression---for the actual spatial field, which fully characterises its probability distribution. 

Deriving actual spatial fields relates closely to the work of \cite{fuentes} in which measurements of particulate matter and simulated data are related through a Gaussian model to the `unobserved truth', for which the posterior distribution is derived. Part of the Extreme Wind Storms (XWS) catalogue \citep{xwspaper} also had the same aim: to combine measurements and simulated footprints to give `recalibrated footprints', which are linear functions of simulated footprints achieved using a mixed effects model. This work may be seen as an extension that derives an actual spatial field using a variant of kriging to relax the assumption of linearity between actual and simulated values. Kriging was used to derive the E-OBS gridded datasets \citep{knmidata, knmidata2} from measurements. Unlike the E-OBS data, we also consider simulated data.

The next section of this paper gives details of the proposed framework for deriving actual spatial fields for environmental processes. We derive actual footprints for European windstorms in Section \ref{eurows} using windstorm Daria for illustration. Section \ref{discuss} summarises the proposed framework.

\section{Proposed framework} \label{meth}

This section describes how we derive actual spatial fields for environmental process using measurements and numerical model simulations.

\subsection{Prior models for spatial fields} \label{prior-haz}

Let $Z(s)$ denote the actual value of some environmental process at location $s$ on region $R$. Let $Y(s)$ denote a measurement on $Z(s)$. We assume that $Y(s) = Z(s) + \epsilon(s)$ where $\epsilon(s)$ are independent measurement errors. We then assume that these errors have zero mean and are Gaussian distributed with variance $\sigma_{Y}^2$, so that \begin{equation} Y(s)\,|\,Z(s) \sim N(Z(s), \sigma_{Y}^2). \label{obsmod}\end{equation}

Now let $X(s)$ denote a simulation of $Z(s)$. (Typically output $X(s)$ will be represented on a grid with the same boundary as $R$. We assume that $X(s)$ can be interpolated onto $R$ for any $s \in R$.) We suppose that \begin{equation} Z(s) = m(X(s)) + \varepsilon(s, X(s)), \label{zgxdecomp}\end{equation} so that an actual value of the process, $Z(s)$, is related to the simulated value, $X(s)$, through a parametric discrepancy term, $m$, and a random discrepancy term, $\varepsilon$, which can vary with space and intensity. We assume a Gaussian process (GP) prior for $Z(s)$ given its simulated counterpart, $X(s)$, which is written \begin{equation} Z(s)\, | \,X(s) \sim GP(m(X(s)), \sigma_X^2 c_X(\, , \,)), \label{cmomod} \end{equation} where $m(\,)$  and $\sigma_X^2 c_X(\, , \,)$ are its mean and covariance functions, respectively. The covariance function will be used to simultaneously allow smooth forms for the spatial discrepancy, $\varepsilon_{spatial}(s)$, and the intensity discrepancy, $\varepsilon_{intensity}(X(s))$ such that $\varepsilon(s, X(s)) = \varepsilon_{spatial}(s) + \varepsilon_{intensity}(X(s))$. 

Relation \eqref{cmomod} represents $[Z(s) \, | \, X(s), \, \Theta]$, where $[ \, ]$ denotes ``distribution of'' and $\Theta$ represents an arbitrary parameter set.  \cite{guttorp} propose a similar decomposition to equation \eqref{zgxdecomp} for $[X(s) \, | \, Z(s), \, \Theta]$. Adopting the prior, $[Z(s) \, | \,  \Theta] \propto 1$, unites our approach with that of \cite{guttorp} for symmetric discrepancy terms, as then $[X(s)\, | \, Z(s), \, \Theta] \propto [Z(s) \, | X(s), \, \Theta]$. For simulation models that act as `smoothers', that is, fail to capture small-scale process, the conditioning direction of relation \eqref{cmomod} offers the interpretation that $Z(s)$ is a function of $X(s)$ plus some noise, as in equation \eqref{zgxdecomp}.

Gaussian formulations for relations \eqref{obsmod} and \eqref{cmomod} are also used in \cite{kenohag}. While these are highly tractable and tend to provide robust modelling choices, as stated in \cite{kenohag} ``equally cogent arguments could probably be evinced in favour of other models''. Transformations to data should also be considered before defining $X(s)$, $Y(s)$ and $Z(s)$. See Section \ref{discuss} for further discussion on this and on relaxing the assumption of unbiasedness, that is, zero-mean measurement errors.

\subsection{The posterior distribution of an actual footprint} \label{recal-fp}

We combine the prior models of relations \eqref{obsmod} and \eqref{cmomod} in Section \ref{prior-haz} to give the marginal model \begin{equation} Y(s)\,|\,X(s), \sigma^2, \beta, \theta \sim GP(m(X(s)), \sigma^2 c(\, , \,)), \label{margmod} \end{equation} where $m(\,)$ is as in Relation \eqref{cmomod} and $\sigma^2 c(\, , \,) = \sigma_Y^2 c_Y(\, , \,) + \sigma_X^2 c_X(\, , \,)$, which is based on writing relation \eqref{obsmod} as a GP with covariance function $\sigma_Y^2 c_Y(\, , \,)$. For tractability, we assume $m(x) = h^T(x) \beta$, where $h(\,)$ comprises $q$ basis functions (eg. $h(x) = (1, x)^T$) and $\beta$ comprises $q$ regression coefficients. Depending on the forms chosen for the correlation functions, not all their parameters, collectively denoted $\theta$, may be identifiable without prior knowledge, in particular if both $c_X(\, , \,)$ and $c_Y(\, , \,)$ contain nugget terms. We address this in Section \ref{eurows} by incorporating prior knowledge on measurement error. We also choose a common correlation function across events. This choice gives stable $\theta$ estimates; see Section \ref{discuss} for further discussion.

The conjugate prior form for $(\beta, \sigma^2)$ is normal inverse-gamma of the form \[\pi(\beta, \sigma^2) \propto (\sigma^2)^{-(d + q + 2)/2} \exp[-\{(\beta - b)^TB^{-1}(\beta - b) + a\}/(2 \sigma^2)],\] where $a$, $b$, $B$ and $d$ are hyperparameters. Consider data on hazard events $j=1, \ldots, J$. Let $D = \{D_1, \ldots, D_J\}$ where $D_j=\{(x_j(s_{k}), y_j(s_{k}))\}_{k=1, \ldots, K_j}$ comprise $K_j$ simulator output and measurement pairs at locations $s_1, \ldots, s_{K_j}$ and $y_j = (y_j(s_{1}), \ldots, y_j(s_{K}))^T$.  As no conjugate prior form for $\pi(\theta)$ exists, and $\theta$ is difficult to elicit, we set $\pi(\theta) \propto 1$ and choose $\theta$ to maximise its posterior distribution \begin{align} &\pi(\theta\, | \, D) \propto \pi(\theta) \pi(D \, | \, \theta) \propto \pi(\theta) \prod_{j=1}^J \hat \sigma_j^{-K_j} |A_j|^{-1/2}, \label{postmode} \\   \shortintertext{where} \hat \beta_j &= B_j^*(B^{-1}b + H_j^T A_j^{-1} H_j)^{-1} H_j^T A_j^{-1} y_j, \nonumber \\ \hat \sigma_j^2 &= (K_j + d)^{-1} (a + b^T B^{-1} b + y_j^T A_j^{-1} y_j + \hat \beta^T(B_j^*)^{-1} \hat \beta) - \sigma_Y^2, \nonumber \\ t_j(x(s)) &= (c_X(x(s), x_j(s_{1})), \ldots, c_X(x(s), x_j(s_{K})))^T, \nonumber \end{align} $H_j$ is a $K_j \times q$ matrix with $k$th row $h^T(x_j(s_{k}))$, $\Sigma_j$ is a $K_j \times K_j$ matrix with $(k, l)$th element $c_X(x_j(s_k), x_j(s_l))$ for $k, l = 1, \ldots K_j$,  $A_j = \Sigma_j + \lambda^2 I_{K_j},$ $\lambda^2 = \sigma_Y^2 / \sigma_X^2$ and $B_j^* = (B^{-1} + H_j^T A_j^{-1} H_j)^{-1}$.

Finally, we obtain an analytical expression for the probability distribution of the actual spatial field for event $j$ at location $s$ given simulated data $X(s)$, which is given by \begin{align} Z_j(s)\, | \,X(s)&, \,\theta, \, D \sim GP(m_j^*(X(s)), \hat \sigma_j^2 c_j^*(X(s), \,)), \label{approxtproc} \\ \shortintertext{with} m^*_j(X(s)) &= h^T(X(s)) \hat \beta_j - \sigma_Y^{-2} t_j(X(s)) (y_j - H_j\hat \beta_j), \nonumber \\ c^*_j(X(s), X(s')) &= c_X(X(s), X(s')) - t_j(X(s)) \Sigma_j^{-1} t_j(X(s')).  \nonumber \end{align} Note that the exact form for relation \eqref{approxtproc} is  Student-$t$ process on $K_j-q$ degrees of freedom, which can be parameterised to have negligible difference from a Gaussian process with corresponding mean and covariance functions for large $K_j - q$. We will typically have large $K_j - q$; hence the form adopted in relation \eqref{approxtproc}.

\subsection{Model checking} \label{mod-check}

The model will be first checked using a semivariogram to compare model-based and empirical (semi)variances, ie. $\sigma^2 c(x(s_{k}), x(s_{k'}))$ and $var(y(s_{k}), y(s_{k'}))$. As multiple variables may be used to define $c(\, , \,)$, this precludes conventional use of a semivariogram. For example, $c(\, ,\,)$ is not entirely spatial and nor is its spatial part isotropic. Details to enable semivariogram use are given in Appendix \ref{emulatordiag}. The fitted model is also checked as a whole, using diagnostic tests developed for GP emulators; see \cite{bas-ohag}. Appendix \ref{emulatordiag} gives further details.



\section{Extreme European windstorm footprints} \label{eurows}

In this section we derive posterior distributions of actual footprints for the 50 extreme European windstorms in the XWS windstorm catalogue \cite[see also \url{http://www.EuropeanWindstorms.org}]{xwspaper}. 

\subsection{Data} \label{data}

\begin{figure}[t]
\begin{center}
\includegraphics[width=0.6\textwidth]{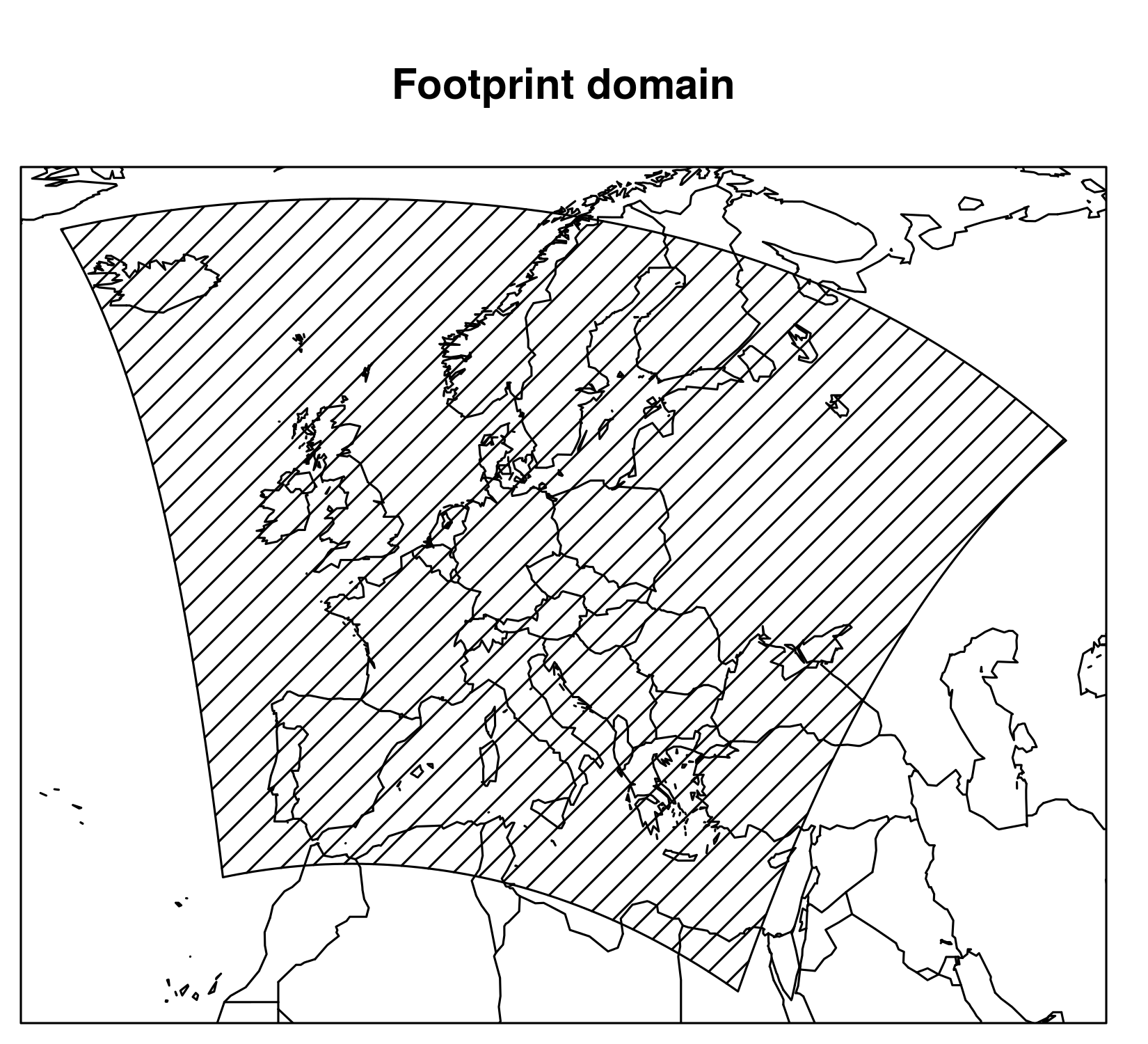}
\caption{\label{domain}Domain of rotated grid on original latitude-longitude scale.}
\end{center}
\end{figure}

The footprint of a windstorm is based on maximum 3 second 10 metre wind gust speeds. For event $j$, $j=1, \ldots, J$, where $J=50$, at location $s$, it is the maximum gust speed attained over a 72-hour period centred on the event's occurrence time. The 50 European windstorms studied have been classed as `extreme' based on their insured loss or meteorological characteristics, although often both coincide. Measurement data and simulated climate model output are both available.  The simulated footprints are generated from ERA Interim re-analysis \citep{erainterim} dynamically downscaled from a horizontal resolution of ~$0.7^\circ$ to $0.22^\circ$ by the Met Office Unified Model \citep{metum}. The footprints are given on a rotated grid. Figure \ref{domain} shows the footprint boundary on original latitude-longitude scale, which is the domain studied here. Measurement data are daily maximum gust speeds taken from the NOAA Global Summary of the Day (GSOD) database\footnote{\url{http://www7.ncdc.noaa.gov/CDO/cdoselect.cmd}}, which have been supplemented with data from the BADC Midas database \citep{midas}.

\begin{figure}[t]
\begin{center}
\includegraphics[width=0.7\textwidth]{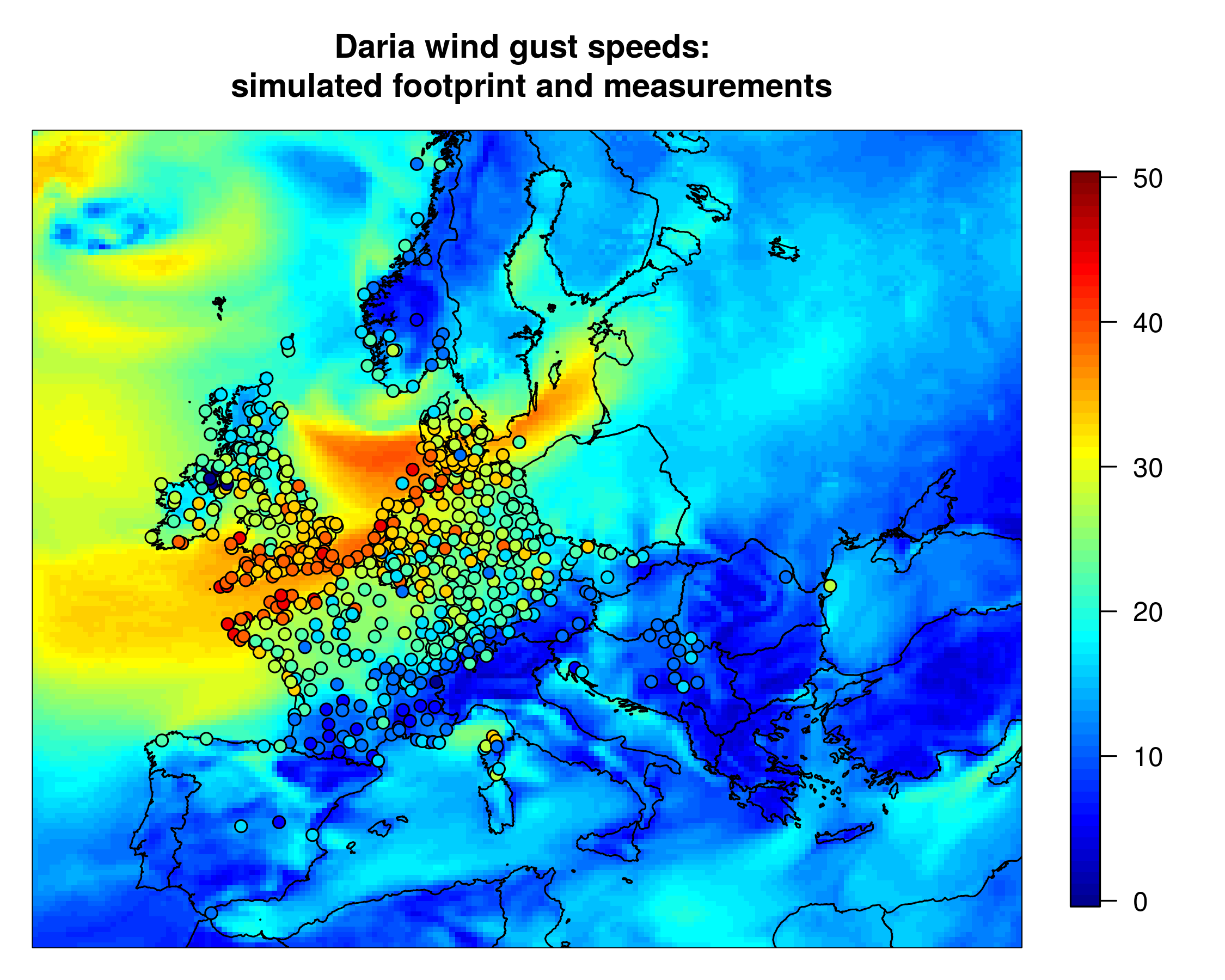}
\caption{\label{daria1}Climate-model-simulated footprint for windstorm Daria together with measured maximum wind gust speeds. Wind gust speeds have units ms$^\text{-1}$.}
\end{center}
\end{figure}

\begin{figure}[h!]
\begin{center}
\includegraphics[width=0.6\textwidth]{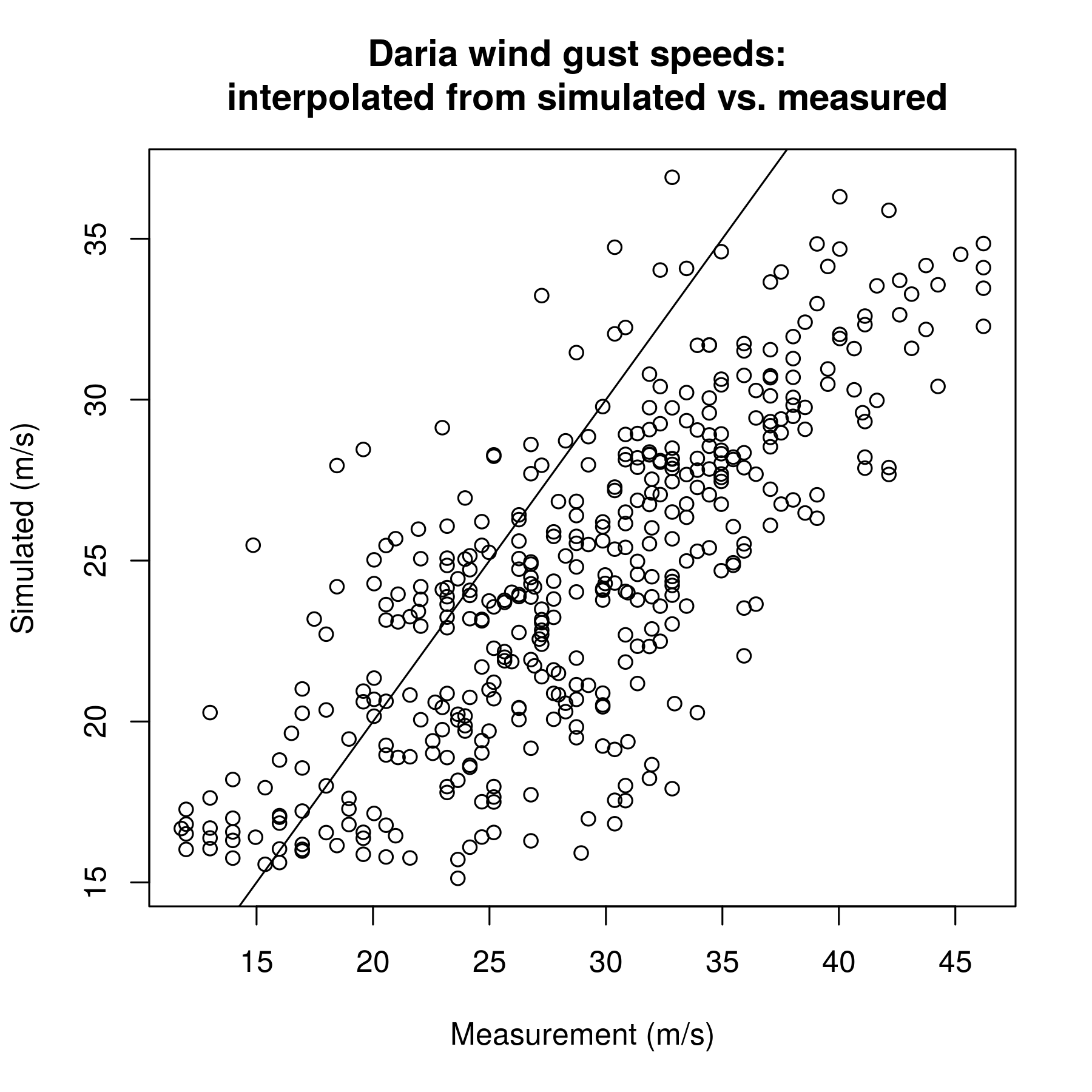}
\caption{\label{daria2}Gust speeds interpolated from simulated Daria footprint against measurements. Line represents $y=x$, ie. perfect agreement.}
\end{center}
\end{figure}

Windstorm Daria (sometimes referred to as The Burns' Day Storm) is used as a case study. Daria reached its peak intensity on 25th January 1990 and had an insured loss estimated at \$8.2bn (indexed to 2012 prices). Figure \ref{daria1} shows Daria's simulated footprint on the rotated grid alongside GSOD measurements. Figure \ref{daria2} shows measurements against wind gust speeds (bi-linearly interpolated) from the footprint, which will be referred to as `interpolated gust speeds'. The interpolated gust speeds clearly do not match the measurements. It is difficult, however, to distinguish from Figure \ref{daria1} regions where agreement is better or worse. The wind gust speeds in the simulated footprint tend to differ more from measurements exceeding 25ms$^\text{-1}$, and appear to asymptote near 30ms$^\text{-1}$, unlike the measurements. This general `bias' in the relationship between the simulated and measured gust speeds is evidence of intensity discrepancy. Although causes of discrepancy in the simulated gust speeds are beyond the scope of this paper, note that discrepancy tends to be greatest for the highest gust speeds, which lead to most damage and loss.

\subsection{Model specification} \label{mod-spec}

Losses very rarely occur if wind gust speeds are below 25ms$^\text{-1}$. Consequently the model proposed in Section \ref{prior-haz} is fitted only to measurement and simulation pairs where the simulated wind gust speeds exceed a threshold, denoted $u$ms$^{-1}$. This allows for more accurate estimation of their relationship for high speeds, which are of greatest interest for loss estimation. Gust speed measurements are also sometimes not recorded if they fail to exceed a certain value; focussing only on high gust speeds helps avoid effects of such selection bias. We choose $u=15$ms$^\text{-1}$, which is judged in part from the relationship between the measured and interpolated gust speeds in Figure \ref{daria2}. This is below the 25ms$^{-1}$ usually needed for losses to occur. 

Imposing the threshold means that we do not use the entire simulated footprint to derive the actual footprint. However, we present the actual footprint for the entire domain, because footprints tend to be simpler to grasp if shown for a complete region, grids are readily accommodated in many software packages, and the censored footprint---corresponding to where the simulated footprint is below the threshold---can be derived from the complete footprint. We choose the mean function for event $j$ to be $h^T(x) \beta_j$, where $h^T(x) = (1, x, x^2)$, with prior probability \[\pi(\beta_j, \sigma_j^2) \propto \sigma_j^{-(q + 2)/2} \exp[-(\beta_j - b)^TB^{-1}(\beta_j - b)/(2 \sigma_j^2)],\]  where $b=(0, 1, 0)^T$ and $B=diag(0.1, 1, 1)$. Our prior is relatively informative only for the intercept term, and reflects our decision to impose a threshold. It is designed to counteract effects of missing measurements, which occur more frequently amongst lower gust speeds. For example, towards the beginning of the measurement period, often gust speeds were only recorded if they exceeded a threshold. The chosen prior and threshold overcomes issues associated with this sampling bias by letting the posterior mean of the actual footprint revert to the simulated footprint at lower gust speeds, which tend to be simulated more accurately, and tend not to be of importance for loss estimation.

%
%

The marginal model is summarised as \begin{equation} Y_j(s)\,|\,X_j(s) > u, \, \sigma_j^2, \, \beta_j, \, \theta \sim GP(h^T(X_j(s)) \beta_j, \sigma_j^2 c(\, , \,)), \label{margmod2} \end{equation} where $\sigma_j^2$ is the event-specific GP variance. Note that the Gaussian assumption applies to $Y_j(s) - h^T(X_j(s))$, thus to a difference that may be considered a residual. This assumption does not impose that wind gust speeds are non-negative. The model will be inappropriate if, given $h^T(X_j(s))$, residuals frequently coincide with negative wind gust speeds. This is found not to be the case here. 


The statistical model is defined on a transformed space: standard latitude-longitude coordinates are adjusted to the rotated grid on which distances between grid cells are more uniform; see Figure \ref{domain}. Locations on the rotated grid, $s^* = (s_{lon}, s_{lat})^T$, are then adjusted to a \emph{transformed space}, with location $s$ defined as $s = (s_1, s_2)^T = T s^*$, where \[T = \left(\begin{array}{rr} \cos \omega & -\sin \omega \\ \sin \omega  & \cos \omega\end{array}\right).\] The unscaled covariance function, $c(\, , \,)$ in relation \eqref{margmod2} (see also equation \eqref{corform} of Appendix \ref{variogdiag}), is chosen have the form \begin{align} c(X_j(s), X_j(s')) &= \left\{\begin{array}{ll} 0 & \text{if}~~j \neq j',\\  1 + \lambda^2 & \text{if}~~s=s', j=j',\\  c_{spatial}(s, s') c_{intensity}(X_j(s), X_j(s')) & \text{otherwise,}\label{corform2}\end{array}\right. \intertext{where $\lambda^2$ is a nugget parameter,} c_{spatial}(s, s') &= \left\{\prod_{k=1}^2 \dfrac{1}{\Gamma(\nu_k)2^{\nu_k-1}}\left(\dfrac{\sqrt{2\nu_k}h_k}{\phi_k}\right)^{\nu_k} K_{\nu_k}\left(\dfrac{\sqrt{2\nu_k}h_k}{\phi_k}\right)\right\}, \end{align} with $\Gamma(\,)$ the gamma function, $K_{\nu_k}(\,)$ the modified Bessel function of the second kind, $h_k = |s_k - s_k'|$, range and smoothness parameters $\phi_k$ and $\nu_k$, $k=1, 2$, respectively, and \[c_{intensity}(X_j(s), X_j(s')) = \exp\left\{-\left(\dfrac{X_j(s) - X_j(s')}{\phi_X}\right)^2\right\},\] with $\phi_X$ a further range parameter. Thus from relation \eqref{margmod2} $\theta=(\omega, \lambda^2, \phi_1, \phi_2, \nu_1, \nu_2, \phi_X)$. The choice of unscaled covariance function in equation \eqref{corform2} follows from partitioning discrepancy into spatial and intensity parts. The spatial discrepancy is then separated into (transformed) longitude and latitude directions. In each direction a different Mat\'{e}rn form is assumed, which reflects prevailing directions of winds in windstorms. Non-zero $\omega$ allows directions non-parallel to the longitude and latitude axes, which, as the prevailing wind direction is approximately south westerly, should be anticipated. A Gaussian covariance form captures intensity discrepancy, as this is expected to be very smooth, which is the main motivation for its regular use in Gaussian process emulators. 

The nugget, $\lambda^2$, represents a nugget in $c_Y(\, , \,)$, which captures measurement error, and a nugget in $c_X(\, , \,)$, which captures small-scale fluctuations in the actual footprint not captured by the spatially-aggregated simulated footprint. The posterior mode for $\lambda$ is approximately 3.5ms$^{-1}$, which puts an upper bound on the measurement error's standard deviation. Consequently we set $\sigma_Y=3\text{ms}^\text{-1}$. This choice is intended to be fairly conservative, yet consistent with simple analysis of the measurements.

We fit the model by finding the posterior mode of $\pi(\theta \, | \, D)$ (equation \eqref{postmode}) numerically.

\subsection{Model checks}

\begin{figure}[t]
\begin{center}
\includegraphics[width=0.32\textwidth]{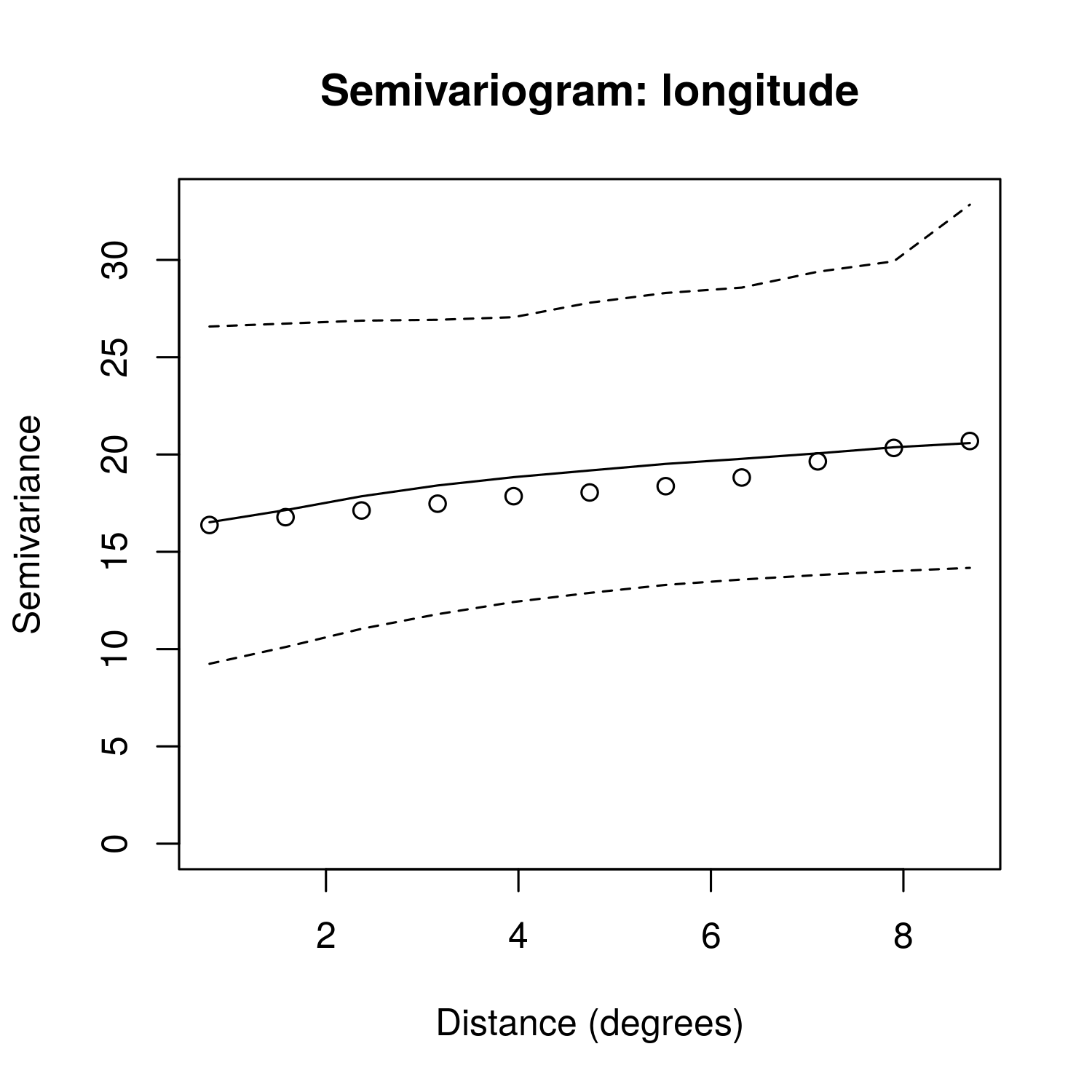} \includegraphics[width=0.32\textwidth]{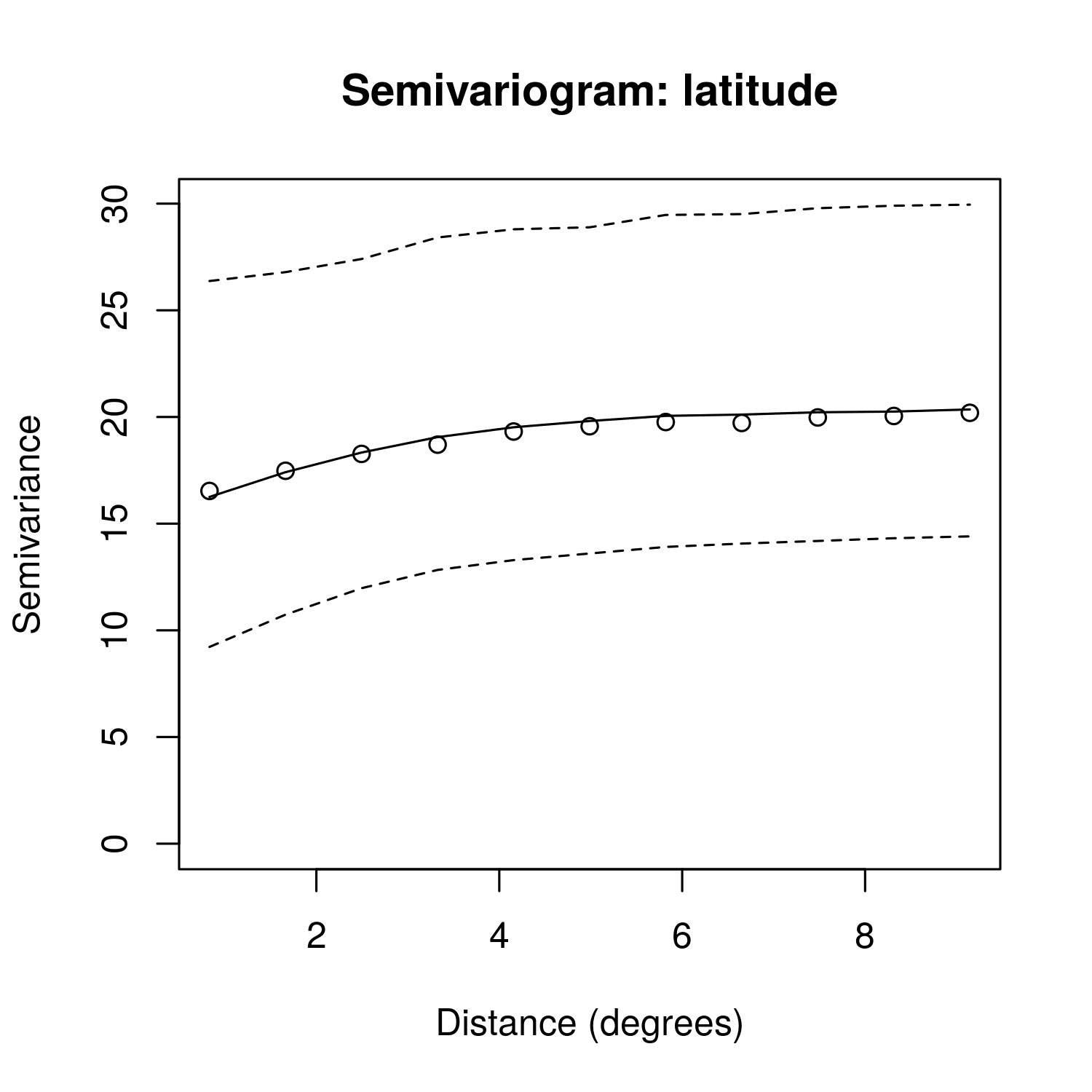} \includegraphics[width=0.32\textwidth]{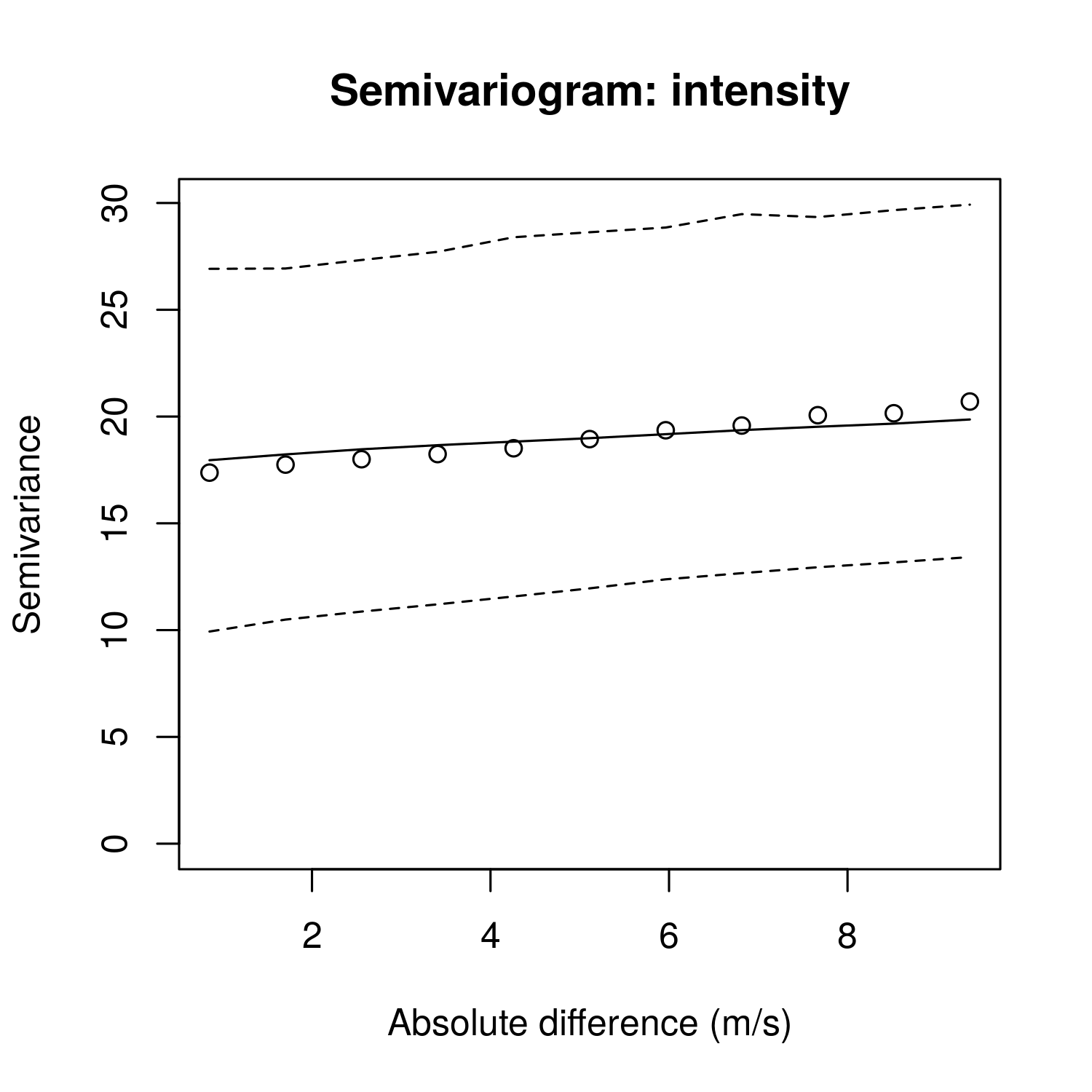}
\caption{\label{variogs}Empirical variograms and model-based estimates for anisotropic semivariances in longitude (left) and latitude (right) directions, for rotated coordinates. Empirical estimates ($\circ$) are shown alongside mean (----) model-based estimates with 95\% uncertainty bounds (- - -).}
\end{center}
\end{figure}

Semivariograms for the fitted marginal model are shown in Figure \ref{variogs}. Good agreement between model-based and empirical estimates of dependence for each discrepancy type can be seen. The spatial dependence structure matches the Mat\'{e}rn form in both transformed directions, while the intensity discrepancy matches the Gaussian form. The bounds on the plots represent 95\% error bounds, estimated from within-bin sampling variability, but do no take into account parametric uncertainty, which is likely to be comparatively small. As the empirical estimates fall well within the error bounds, this offers very good support for that fit of the marginal model, which relates measured wind gust speeds to those in the simulated footprint. This is confirmed by similar findings for the remaining 49 windstorms (not shown).

The validation plots of Figure \ref{valid} use the procedures described in Appendix \ref{emulatordiag} to check the model's assumptions simultaneously. 30 pairs of measured and simulated footprint wind gust speeds are withheld as validation data; the measurements are then compared against their predictive distributions, given the 30 simulated footprint gust speeds. The four plots of Figure \ref{valid} support the predictions for windstorm Daria: the standardised residuals and pivoted Cholesky errors are consistent with their respective $t$ distributions, and the Mahalanobis distance with the $F_{\tilde n, n_j - q}$ distribution. These show the measurements to be consistent with their predictive distribution. Similar results are found for the remaining windstorms (not shown). Coupled with the semivariogram check, we conclude that there is good agreement between the statistical model and the validation data.

\begin{figure}[t]
\begin{center}
\includegraphics[width=0.4\textwidth]{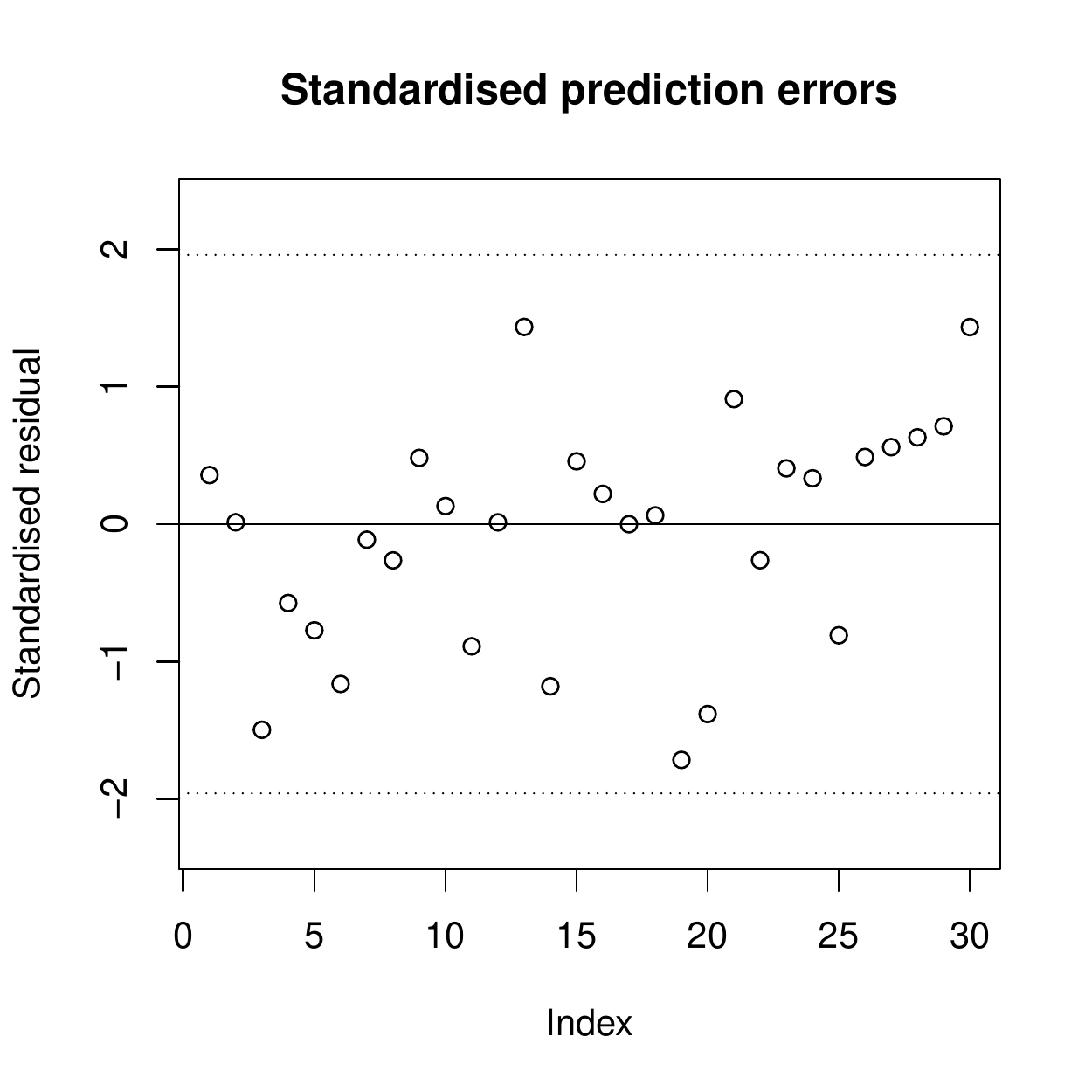} \includegraphics[width=0.4\textwidth, page=2]{valid} \includegraphics[width=0.4\textwidth, page=3]{valid} \includegraphics[width=0.4\textwidth, page=4]{valid}
\caption{\label{valid}GP validation plots: see Appendix \ref{emulatordiag}.}
\end{center}
\end{figure}

\subsection{Posterior estimates of actual European windstorm footprints}

\begin{figure}[h!]
\begin{center}
\includegraphics[width=0.48\textwidth]{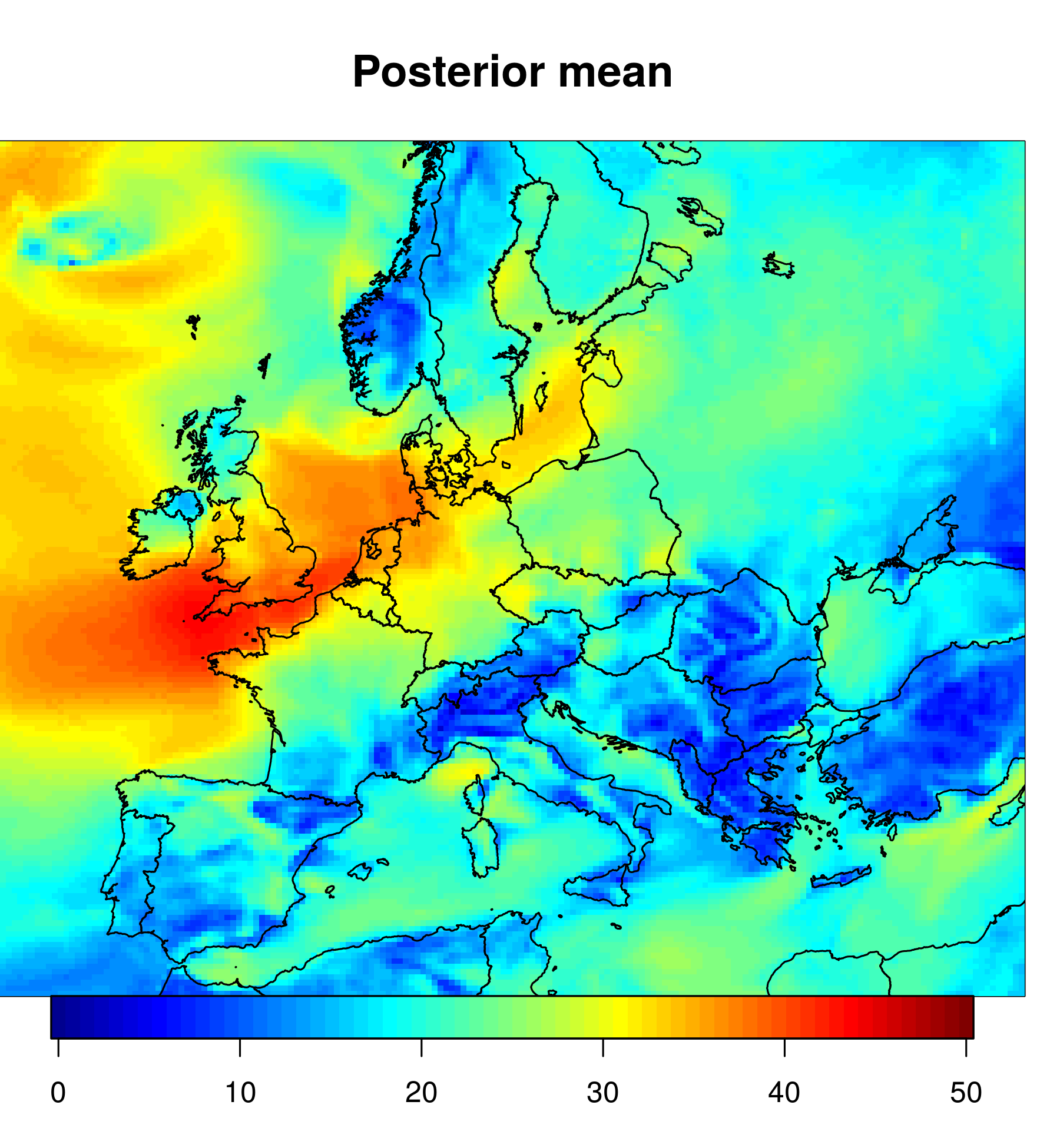} \hfill \includegraphics[width=0.48\textwidth]{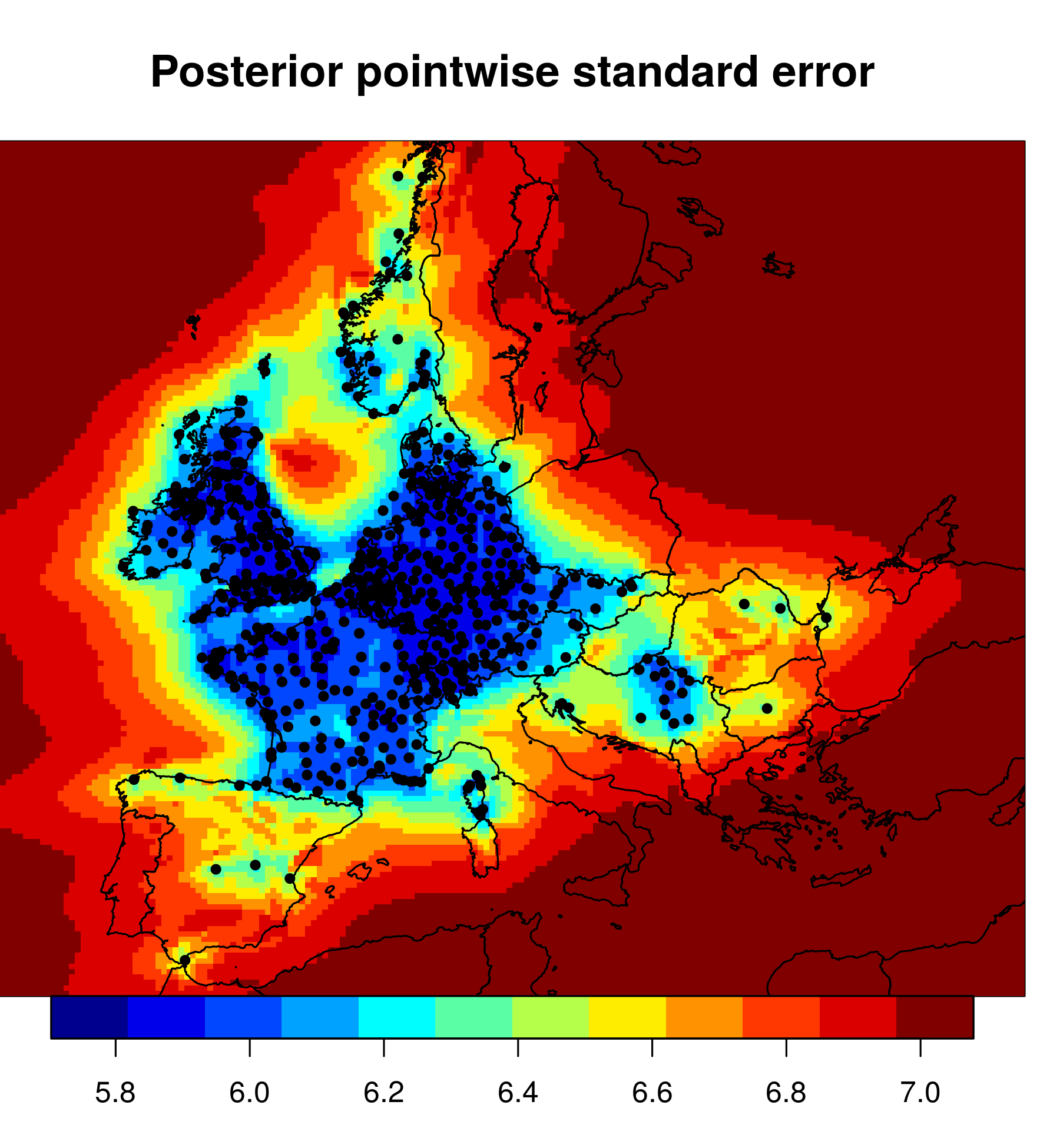} \includegraphics[width=0.48\textwidth]{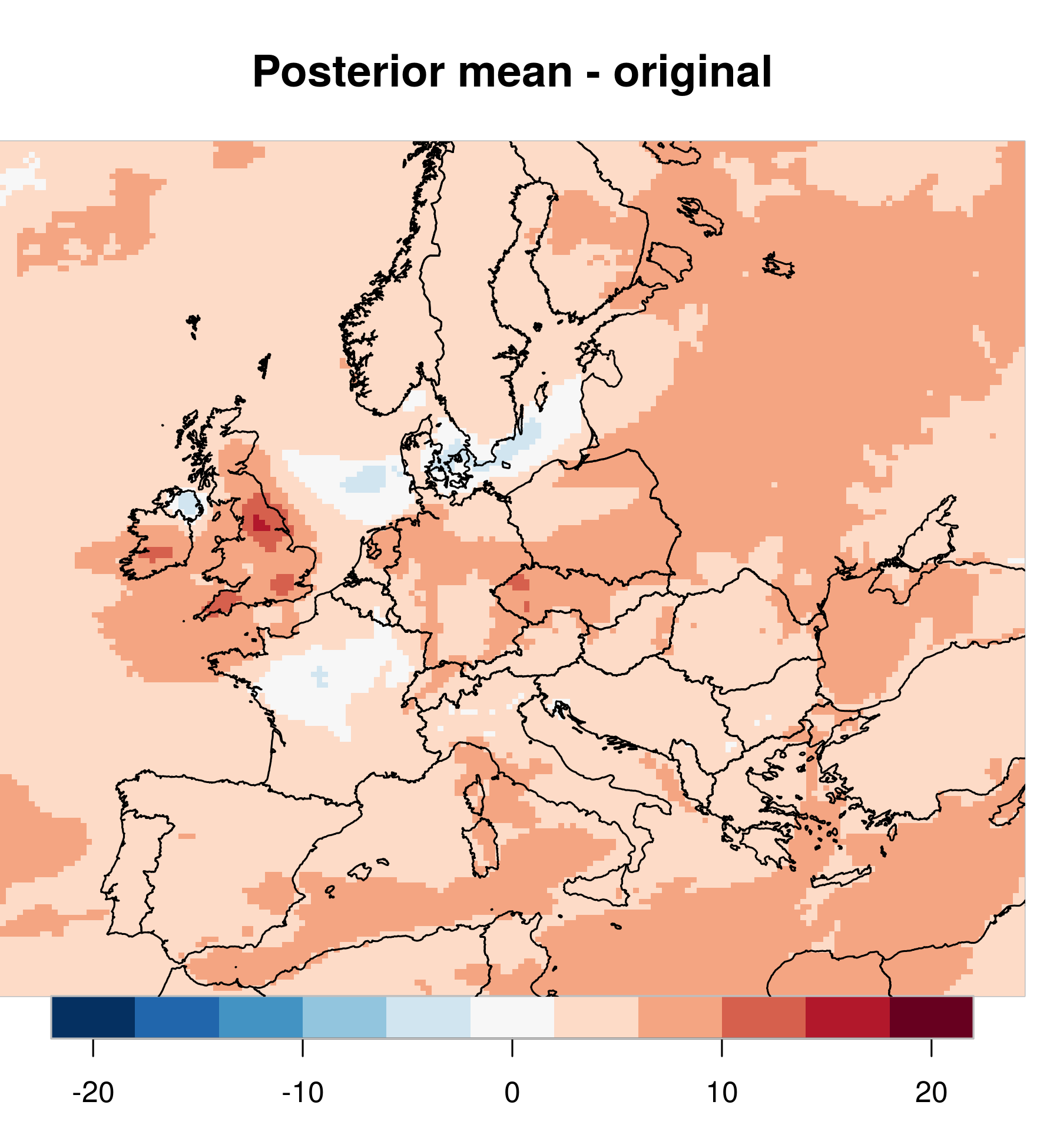} \hfill \includegraphics[width=0.48\textwidth]{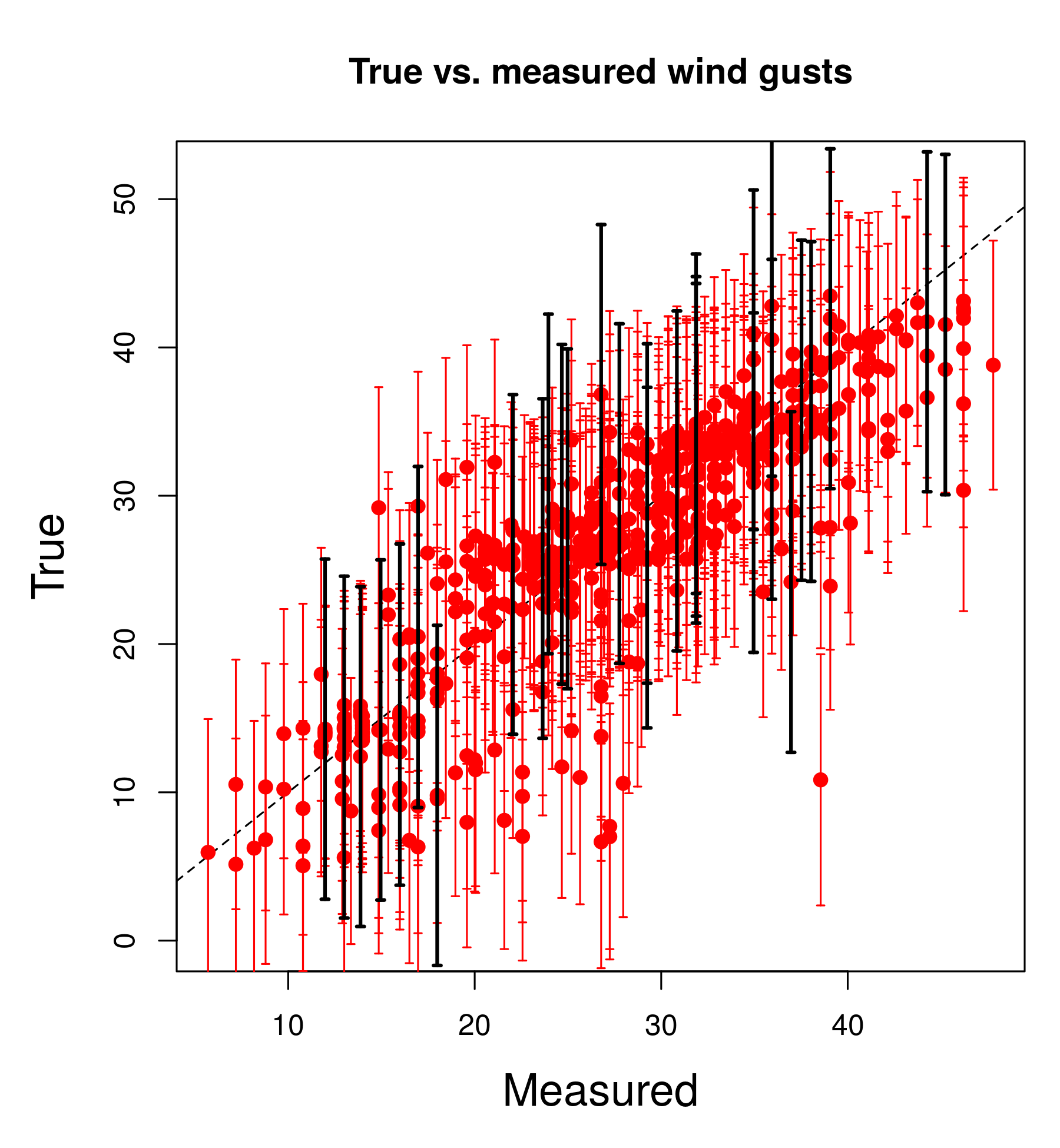}
\caption{\label{recal}Maps representing posterior estimate of the actual footprint for windstorm Daria. Its posterior mean (upper left), pointwise standard errors (upper right) and the difference between posterior mean and original simulated footprint (lower left) are shown. Posterior mean estimates with 95\% prediction intervals of gust speeds for stations used in model fitting (red) and validation stations (black) are plotted against measurements (lower right).}
\end{center}
\end{figure}

The actual footprint for windstorm Daria is represented in Figure \ref{recal} by its posterior mean and standard errors. The posterior mean gives plausible spatial structure for a windstorm footprint, preserving much of the structure in the original climate-model-simulated footprint. Uncertainty in the estimate of the actual footprint is smallest where most measurements are present and increases as measurements become fewer and more distant, which should be expected. 

For windstorm Daria, the root mean squared error (RMSE) between the measured and interpolated gust speeds is $7.69$ms$^{-1}$. This reduces to $4.80$ms$^{-1}$ for the posterior mean of the actual footprint. This is supported by Figure \ref{recal}, which shows that posterior mean estimates for the actual footprint are closer to $y=x$ line than the simulated wind gust speeds (from Figure \ref{daria2}). The predictions for the 30 validation measurements are concentrated around the $y=x$ line. (Note that these uncertainties are larger than for the actual footprint, as they also include measurement error.) The gust speeds in the posterior estimate of the actual footprint are consistently greater than those simulated by the MetUM climate model, which is reflected in the ratio of Figure \ref{recal} exceeding one for most of the region studied. That the actual footprints better match the measurements than the simulated footprints is supported by considering posterior estimates of actual footprints for all 50 windstorms together when calculating the RMSE, which reduces from $5.39$ms$^{-1}$ to $3.62$ms$^{-1}$. 

\section{Discussion} \label{discuss}

We have presented a framework for deriving posterior distributions for actual values of spatial fields representing environmental processes from measurements and simulated data. The method readily incorporates the natural belief that the actual spatial field exhibits some degree of smoothness over space; capturing spatial dependence pools information across locations to give consistent spatial estimates. Gaussian processes are used to model two forms of discrepancy: spatial discrepancy, which is the spatial difference between the two fields; and intensity discrepancy, which is the systematic bias in intensity between simulated and actual values of the spatial process. The result of this model is that, given the measurements and simulated spatial fields, the posterior distribution of the actual spatial field has a closed-form Gaussian process representation, which gives a full characterisation of its uncertainty.

There has been little research into methods for estimating actual spatial fields. Our method may be thought of as an extension of the XWS recalibration method \citep{xwspaper}, which is based on a random effects regression model, that allows for spatial dependence and removes the constraint that the actual footprint is a function of the simulated footprint that is linear in some parameters. The desirable criterion that the relationship between the simulated and actual footprint may vary between windstorms is preserved. Our work could be classed as a post-processing method, which are seen more commonly in the forecast verification literature, with methods such as bias correction or recalibration; see \cite{goddard} for details.




The biggest constraints in the proposed method are perhaps imposed by the Gaussian assumptions; these are present in the measurement error model, and the Gaussian process priors for the spatial and intensity discrepancy. Transformations to the measurements, the simulator output, or both, can be used to make the Gaussian assumptions less restrictive. The Box-Cox class of transformation may prove useful \citep{boxcox}. There is also scope for improvement to the covariance structure. While the mean and variance of discrepancy between actual and simulated spatial fields have been allowed to vary between events, the correlation function has not. This decision helps give stable and reliable estimates of the correlation function. This constraint may not be necessary for applications to spatial processes with more measurements available. A more complex covariance function, in which distance between locations is separated into over-land and over-sea distances, may also benefit the modelling of windstorm events because a windstorm may slow down over land, due to drag, but speed up over sea, where drag is less; see, for example, \citet[Chapter 7]{wallhobb}. Extensions to the GP's mean function, such as including a covariate representing distance to the coast, perhaps in the prevailing wind direction, may also bring a more accurate model for windstorms.

Much or our model's simplicity hinges on the marginal formulation of relation \eqref{margmod} and conjugate priors. This allows all but the correlation function parameters to be integrated out. Although we have used an informative prior for $\beta$, there is scope for further prior information to be incorporated, in particular to relax the assumption that the measurements are unbiased. Without prior information, assuming linear forms for bias, for example, could cause identifiability problems with the intensity discrepancy. Here we do not have sufficient evidence to suggest bias in measurements. 

\begin{figure}[t]
\begin{center}
\includegraphics[width=.98\textwidth]{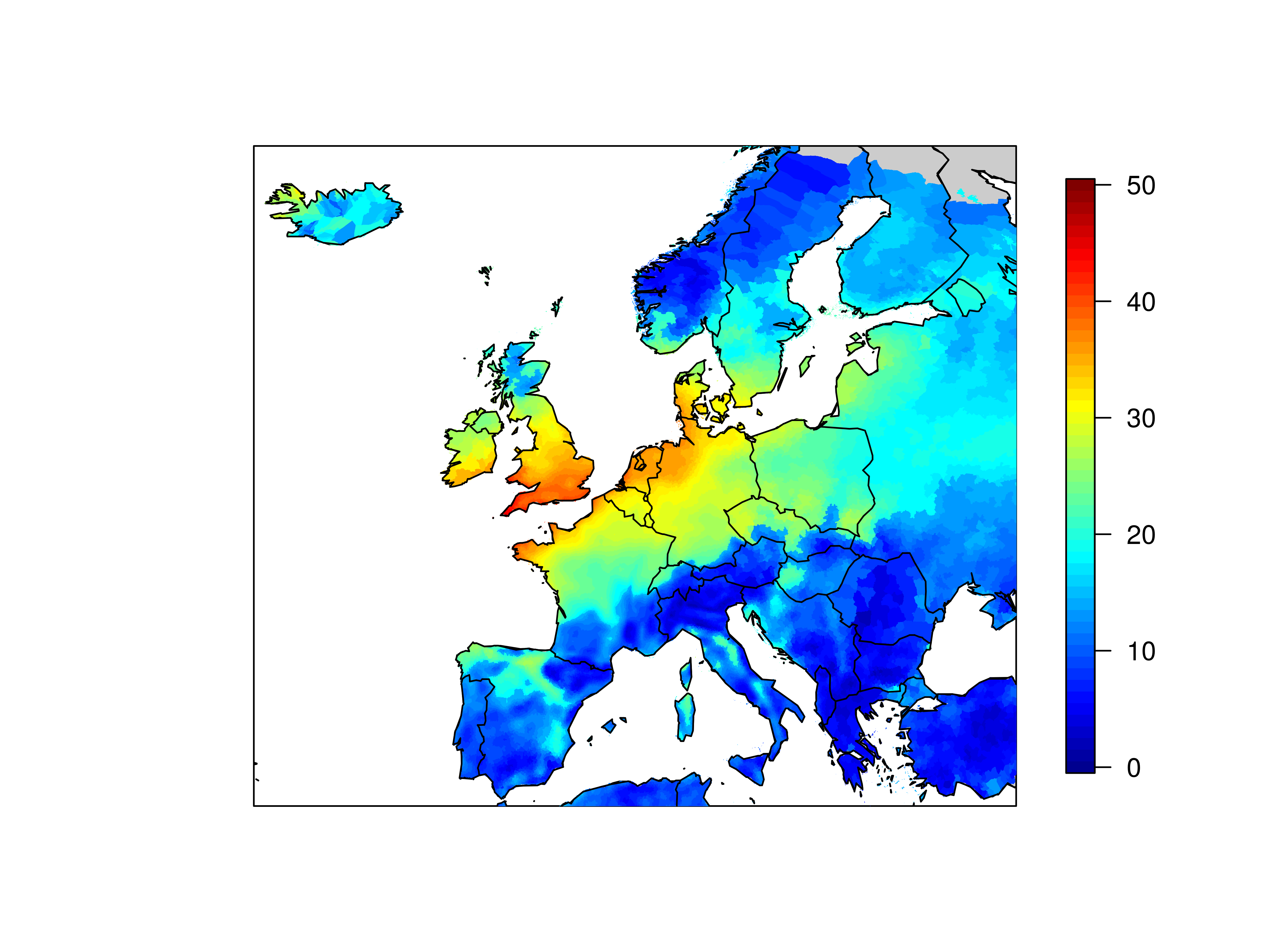} 
\caption{\label{postcode}Posterior mean estimate of actual footprint for windstorm Daria based on administrative areas for Europe.}
\end{center}
\end{figure}

Our proposed model has benefits for estimating financial loss. The posterior distribution of actual gust speeds can be derived for irregularly spread locations. Figure \ref{postcode} shows an alternative posterior mean estimate for the actual footprint for Daria based on standard latitude-longitude scale for administrative areas\footnote{Administrative area data downloaded from \url{http://www.gadm.org/}; accessed \today.}; for the UK, for example, these correspond to postcode sectors, such as `EX4 4**'. Note that sizes of administrative areas vary significantly within and between countries. The ability to derive and present footprints on such scales is convenient for the insurance industry for compatibility with exposure data, such as sums insured. This enables loss calculations to be based on actual wind gust speeds, while accounting for uncertainty in their values. 


Finally, realisations of actual spatial fields can also be simulated using model \eqref{approxtproc} to give synthetic `event sets'. These can be constrained to a specified event's measurements and use to give alternative scenarios that are consistent with its measurements while having realistic spatial structure. Such scenarios are sometimes used in risk estimation.

\section*{Acknowledgements}

We thank Phil Sansom for helpful discussion. We thank the Willis Research Network for supporting this work, the Met Office for providing the windstorm measurement data, and Julia Roberts for help with data provision.

\bibliographystyle{chicago}
\bibliography{recalbib}

\appendix

\section{Appendix to Section \ref{mod-check}} \label{appmodvalid}
%
%
%
\subsection{Semi-variograms} \label{variogdiag}

Consider a general form for the correlation function, so that for $s$ and $s'$ \begin{equation} c(s, s') = \prod_{k=1}^{p'} c_k(w_k(s), w_k(s)') + \sum_{k=p'+1}^p \sigma_k^2 c_k(w_k(s), w_k(s)'), \label{corform} \end{equation} where $\{w_k(s)\}$ are variables relating to location $s$. Define $e(s) = y(s) - h^T(s) \beta$. For a given event $var(e(s), e(s'))$ can be plotted against $\hat \sigma_j^2 c(s, s')$ for all pairwise combinations of $s$ and $s'$. Note that where there are multiple events, combinations across events are not considered, as independence is assumed between events.

To produce a more stable and informative comparison of $var(e(s), e(s'))$ and $c(s, s')$, binning, as is common in the geostatistics literature, is used. Both $var(e(s), e(s'))$ and $c(s, s')$ are binned, according to bins defined by any variable $w_k(s)$, and then plotted against the bins defined by the chosen variable. This is a general and informative way of checking agreement between the model-based covariance structure and the dependence exhibited by observations on the natural hazard event.

\subsection{Emulator diagnostics} \label{emulatordiag}


These diagnostics summarise some of those presented in \cite{bas-ohag}. For event $j$ consider $\tilde n$ validation points $\tilde Y = (Y(\tilde s_{1}), \ldots, Y(\tilde s_{\tilde n}))^T$ at locations $\tilde s_{1}, \ldots, \tilde s_{\tilde n}$ with simulator output $X(\tilde s_{1}), \ldots, X(\tilde s_{\tilde n})$. Define matrix $\tilde V$ to have $(k, l)$th element $\hat \sigma_j^2 c^\dag(X(\tilde s_{k}), X(\tilde s_{l}))$, for $k, l= 1, \ldots, \tilde n$ and the vector $\tilde m = (m_j^\dag(X(\tilde s_{1})), \ldots, m_j^\dag(X(\tilde s_{\tilde n})))'$, where \begin{align*} m^\dag_j(X(s)) &= h^T(X(s)) \hat \beta_j - t_j(X(s)) (y_j - H_j\hat \beta_j),\\ c^\dag_j(X(s), X(s')) &= c(X(s), X(s')) - t_j(X(s)) A_j^{-1} t_j(X(s')).\end{align*} These alternative GP mean and covariance functions account for the fact that the measurements that form the validation data are subject to measurement error, unlike the actual values. It follows from model \eqref{margmod} that $\tilde Y \sim MVN_{\tilde n}(\tilde m, \tilde V)$.

Now define $\tilde e(s_{i})= (Y(\tilde s_{i}) - m^\dag(X(\tilde s_{i})))/\sqrt{\hat \sigma_j^2 c^\dag(X(s_{i}), X(s_{i}))}$ for $i=1, \ldots, \tilde n$. Test 1 is based on the approximate distributional result that $\tilde e(s_{i}) \sim N(0, 1)$; thus large $\tilde e(s_{i})$ relative to the standard Gaussian distribution highlight model inadequacy. In the present case, plotting $\tilde e(s_{i})$ against $i$ or spatially against $\tilde s_{i}$ may help show regions in which the predictions are poor. Now consider the pivoted Cholesky decomposition of $\tilde V$ so that permutation matrix $P$ and upper triangular matrix $U$ satisfy $P^T \tilde V P= U^TU$.  Define $G=PU^T$. Elements of the vector $e^{PC}=G^{-1}(\tilde Y - \tilde m)$, where $e^{PC} = (e_1^{PC}, \ldots, e_{\tilde n}^{PC})^T$, are independent and approximately satisfy $e_i^{PC} \sim N(0, 1)$. Test 2 is a quantile-quantile plot of these errors and Test 3 plots $e_i^{PC}$ against $i$. Model inadequacy is indicated in Test 2 with points deviating from the line with intercept zero and unit slope, and in Test 3 with large or non-random values. Test 4 uses the Mahalanobis distance $D_{MH}= \tilde e_i^T \tilde e_i$, for which $D_{MH} \sim F_{\tilde n, n-q}$, the F-Snedecor distribution with degrees of freedom $\tilde n$ and $n - q$. Large $D_{MH}$ relative to $F_{\tilde n, n-q}$ indicates model inadequacy.

\end{document}